\documentclass[preprint,fleqn,showpacs,showkeys, nodayofweek]{revtex4}

\newcommand{\alp}{\alpha}

\newcommand{\omg}{\omega_{\rm BD}}

\newcommand{\tha}{\theta}

\newcommand{\half}{\frac{1}{2}}

\newcommand{\nab}{\nabla}

\newcommand{\beqn}{\begin{eqnarray}}
\newcommand{\eeqn}{\end{eqnarray}}
\newcommand{\be}{\begin{equation}}
\newcommand{\ee}{\end{equation}}

\usepackage{graphicx}
\usepackage{amssymb}
\usepackage{amsmath}
\usepackage{color}
\usepackage{bm}
\usepackage{datetime}
\begin{document}
\title{Equations of State in the  Brans-Dicke cosmology}

\author{Hyung Won Lee}
\email{hwlee@inje.ac.kr}

\author{Kyoung Yee Kim}
\email{kimky@inje.ac.kr}

\author{Yun Soo Myung}
\email{ysmyung@inje.ac.kr}

\affiliation{Institute of Basic Science and School of
Computer Aided Science, Inje University, Gimhae 621-749, Korea}

\begin{abstract}
We investigate  the Brans-Dicke (BD) theory with the potential as
cosmological model to explain the present accelerating universe. In
this work, we consider the BD field as a perfect fluid with the
energy density and pressure  in the Jordan frame. Introducing the
power-law potential and the interaction with the cold dark matter,
we obtain the phantom divide which is confirmed by the native and
effective equation of state.  Also we can describe the metric $f(R)$
gravity with  an appropriate potential, which shows a future
crossing of phantom divide in viable $f(R)$ gravity models when
employing the native and effective equations of state.
\end{abstract}


 \shortdate
\maketitle

\section{Introduction}

Supernova (SUN Ia) observations has shown  that our universe is
accelerating~\cite{SN}. Also cosmic microwave background
radiation~\cite{Wmap}, large scale structure~\cite{lss}, and weak
lensing~\cite{wl} have indicated that the universe has been
undergoing an accelerating phase since the recent past.   Although
there exist a number of models explaining an accelerating universe,
the two promising candidates are the dark energy of cosmological
constant in the general relativity~\cite{cst} and  a modified
gravitational theory such as $f(R)$ gravity~\cite{NO,sf,NOR}.

Recently, there was an extensive study of dark energy models based
on the Brans-Dicke (BD) theory interacting with the cold dark matter
(CDM)~\cite{FT}.   However, the total equation of state $w_{\rm
eff}=p_{\rm tot}/\rho_{\rm tot}$ was mainly used to measure the
evolution of the universe. In order to see how the BD field
describes the accelerating phase, we have to introduce both the
native and effective equations of state because the BD field is
non-minimally coupled to gravity~\cite{KimHS,Kim05}. Especially, we
need the effective equation of state to take into account the
universe evolution properly because there always exists an
interaction between the BD fluid and the CDM~\cite{ZP1}. For the
interacting holographic dark energy models, there is no phantom
phase when using the effective equation of state~\cite{KLM} instead
of the native equation of state~\cite{WGA}. For the brane
interacting holographic dark energy models, the effective equation
of state was used to account the evolution of the
universe~\cite{KLMb}.   The effective equation of state could read
off from the Bianchi identity which provides a non-standard
conservation law.

On the other hand, $f(R)$  gravity models  have been extensively
employed to explain the present accelerating universe. The
observational data might imply the crossing of the phantom divide
$W_{DE}=-1$ in the near past~\cite{pdiv}. In this case,  the
crossing of the phantom divide could be resolved  in the viable
$f(R)$ gravity models~\cite{ANO,Bamba,f-models}. Especially, we
would like to mention that a general approach to phantom divide in
$f(R)$ gravity was investigated in~\cite{Bamba}, where the
scalar-tensor version of $f(R)$ gravity was used to see the phantom
divide. However, it was shown that any singular $f(R)$ gravity may
be done non-singular~\cite{sinf}. More recently, consistent, viable
and non-singular $f(R)$ gravity was suggested in~\cite{goodf}.

Interestingly, it was shown that the viable four $f(R)$ models
generally exhibit the crossing of the phantom divide in the future
evolution~\cite{BGL}.

 A common feature to all
analysis was performed by mapping the Starobinsky model~\cite{Star}
to a scalar-tensor theory of gravity.   It seems that the metric
$f(R)$ gravity is equivalent to the BD theory with $\omg=0$, while
the Palatini $f(R)$ gravity is equivalent to the BD theory with
$\omg=-3/2$~\cite{sf}. Despite its mathematical equivalence, two
theories may have shown physically non-equivalence~\cite{CNOT}:
super-accelerating phase in the BD theory describes decelerating
phase in $f(R)$ gravity.  Also, it was pointed out that the mapping
seems to be problematic because the scalar potential defined by
$U(\Phi(R))=R \Phi-f(R)$ with $\Phi=\partial_R f(R)$ induces a
singularity in the cosmological evolution~\cite{BENO,Frolov,JPS}.

Before we proceed, we wish to mention the  difference between
Einstein and Jordan frames~\cite{FT}. We consider the frame in which
non-relativistic matter (CDM, baryons) obey the standard continuity
equation with $\rho_m \sim a^{-3}$. This is the Jordan frame as the
physical frame in which physical quantities are compared to
observations. It is sometimes useful to introduce the Einstein frame
where a canonical scalar field is coupled to non-relativistic matter
directly~\cite{ENOSF}. Even though one considers the same physics in
both frames, using different time and length scales may offer the
apparent difference between the observables  in two frames.

In this work we investigate how the present accelerating phase is
realized in the scalar-tensor theory (BD cosmology). We consider
{\it the BD field as a perfect fluid}  with the energy density and
pressure  in the Jordan frame.  Introducing the power-law potential
(\ref{p-pot})
 and the interaction with the CDM, we
confirmed  the appearance of  phantom divide by using the native and
effective equation of state. Especially, inspired by the work of
Ref.\cite{BGL}, we study the cosmological implications of the $f(R)$
gravity using the BD theory with an appropriate BD potential
(\ref{potential}), which indicates a future crossing of phantom
divide in viable $f(R)$ gravity models when employing the native and
effective equations of state. This shows a close connection between
BD theory and $f(R)$ gravities for explaining future crossing of
phantom divide.    In the BD approach, we find a singularity in the
past evolution of the universe. Hereafter, we consider the metric
$f(R)$ gravity only and thus, we mean $f(R)$ gravity by the ``metric
$f(R)$" gravity.

\section{ BD cosmology without a potential}

For cosmological purpose, we introduce the  Brans-Dicke (BD) action
with a matter in the Jordan frame \be \label{ACT}
   S = \int d^4 x\sqrt{-g}
       \Big[
             \frac{1}{16\pi G}
             \Big(
                   \Phi R - \omega_{\rm BD} \frac{\nab_{\alp}\Phi\nab^{\alp}\Phi}{\Phi}
             \Big)
            +{\cal L}_m
       \Big],
\ee where $\Phi$ is the BD scalar, $\omega_{\rm BD}$ is the
parameter of BD theory, and ${\cal L}_m$ represents other matter
which takes a perfect fluid form.   The field equations for metric
$g_{\mu\nu}$ and BD scalar $\Phi$ are
\beqn
   &&G_{\mu\nu} \equiv R_{\mu\nu} - \half g_{\mu\nu}R
               =  8\pi G T_{\mu\nu}^{BD} + \frac{8\pi G}{\Phi} T_{\mu\nu}^m, \\
  &&\nab^2\Phi = \frac{8\pi G}{2\omega_{\rm BD} +
   3}{T^{m \alp}~_{\alp}},
\eeqn
where the energy-momentum tensor for the BD scalar is defined
by \be
   T^{\rm BD}_{\mu\nu}=
\frac{1}{8\pi G}\Big[\frac{\omega_{\rm
BD}}{\Phi^2}\Big(\nabla_{\mu}\Phi\nabla_{\nu}
\Phi-\frac{1}{2}g_{\mu\nu}(\nabla
\Phi)^2\Big)+\frac{1}{\Phi}\Big(\nabla_{\mu}\nabla_{\nu}\Phi-g_{\mu\nu}
\nabla^2\Phi\Big)\Big] \ee and the energy-momentum tensor for a
perfect fluid takes the form \be \label{emten}
   T^m_{\mu\nu} = p_m g_{\mu\nu} + (\rho_m + p_m)u_{\mu}u_{\nu}.
\ee $\rho_m~(p_m)$ denote the energy density (pressure) of the
matter and $u_{\mu}$ is a four velocity vector with
$u_{\alp}u^{\alp} = 1$.

 Considering  that our universe is
homogeneous and isotropic, we work with the flat
Friedmann-Robertson-Walker (FRW) spacetime \be
  \label{frw} ds^2 = -dt^2 + a^2(t)
                  \Big[ dr^2
                      + r^2 ( d\tha^2 + \sin^2\tha d\phi^2)
                  \Big]. \quad
\ee In this spacetime, the first Friedmann and BD scalar equations
take the forms \beqn \label{FRID}
   &&H^2 + H \Big( \frac{\dot{\Phi}}{\Phi} \Big)
       - \frac{\omg}{6}\Big( \frac{\dot{\Phi}}{\Phi} \Big)^2
       = \frac{8\pi G}{3}\frac{\rho_m}{\Phi},\\
\label{PHIF}&&   \ddot{\Phi} + 3H \dot{\Phi} = \frac{8\pi G(
\rho_m-3p_m) }{2\omega_{\rm BD} +
   3},
\eeqn where $H = \dot{a}/a$ is the Hubble parameter and the overdot
denotes the derivative with respect to  time $t$. Here we note that
the case of $\omega_{\rm BD}=-3/2$ is not allowed unless a
radiation-matter with $p_m=\rho_m/3$ comes into the BD theory.
Regarding the BD field as a perfect fluid, its energy and pressure
are
  defined  by kinetic terms as \cite{KimHS,Kim05}
  \beqn \label{rhobd}
   \rho_{\rm BD} &=& \frac{1}{16\pi G}
                \Big[
                      \omega_{\rm BD} \Big( \frac{\dot{\Phi}}{\Phi} \Big)^2
                    - 6H \frac{\dot{\Phi}}{\Phi}
                \Big], \\
\label{prebd}   p_{\rm BD}    &=& \frac{1}{16\pi G}
                \Big[
                      \omega_{\rm BD} \Big( \frac{\dot{\Phi}}{\Phi} \Big)^2
                    + 4H \frac{\dot{\Phi}}{\Phi}
                    + 2\frac{\ddot{\Phi}}{\Phi}
                \Big].
\eeqn If one does not specify the parameter $\omega_{\rm BD}$, one
cannot determine the BD equation of state  exactly. However, the
Bianchi identity of $\nab_\mu G^{\mu\nu}=0$ implies that there
exists an energy transfer between BD fluid and matter \be
\label{BDC}
   \dot{\rho}_{\rm BD} + 3H( \rho_{\rm BD} + p_{\rm BD} )
           =\frac{1}{G}\frac{\rho_m}{\Phi}\frac{\dot{\Phi}}{\Phi}.
\ee  This continuity  equation play a crucial role because it shows
manifestly the energy transfer between  $\rho_{\rm BD}$ and
$\rho_{m}$ and, thus, it defines the effective equation of state.

\begin{figure}[t!]
   \centering
   \includegraphics{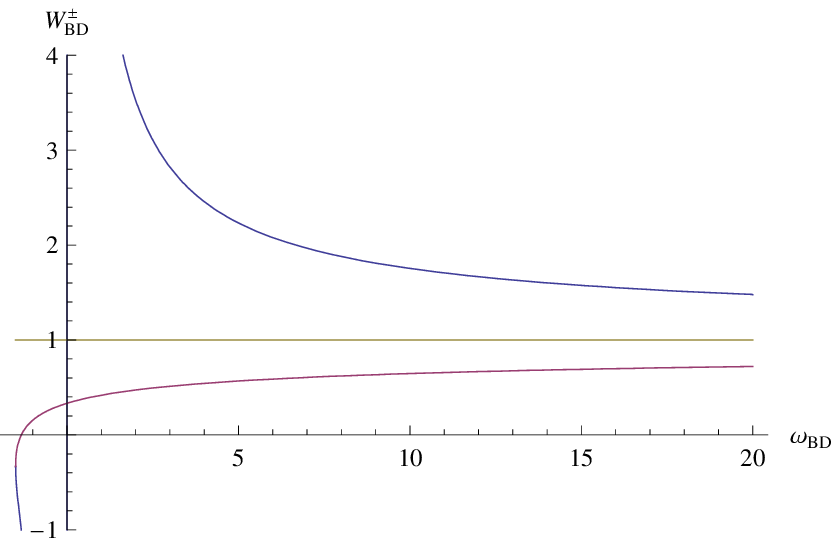}
   \caption{The  equations of state $W^{\pm}_{\rm BD}$ for BD scalar versus its parameter $\omega_{\rm BD}$.
   $W^+_{\rm BD}$ is a monotonically decreasing function of
   $\omega_{\rm BD}$, while $W^-_{\rm BD}$ is a monotonically increasing function of
   $\omega_{\rm BD}$.
    The bound of $W^-_{\rm BD}$ is given by $-1/3 \le W^-_{BD}\le 1$
    because of $\omega_{\rm BD}\ge -3/2$ and $W^-_{BD} \to 1$ as $\omega_{\rm BD} \to
    \infty$. At $\omega_{\rm BD}=0$, one finds that $W^-_{\rm BD}=-\frac{1}{3}$, but $W^+_{\rm BD}$ blows up.}
   \label{eff_w}
\end{figure}

On the other hand,  we consider action with a minimally coupled
scalar $\psi$~\cite{ENOSF}
 \be \label{CACT} \tilde{S} = \int d^4
\tilde{x}\sqrt{-\tilde{g}}
       \Big[
             \frac{1}{16\pi G}\tilde {R}-\frac{1}{2}\tilde{\nab}_{\alp}\psi \tilde{\nab}^{\alp}\psi
   +{\cal L}_m    \Big]
\ee in the Einstein frame. The field equations for metric
$\tilde{g}_{\mu\nu}$ and a scalar $\psi$ are \beqn
   &&G_{\mu\nu}
               =  8\pi G T_{\mu\nu}^{\psi} + 8\pi G T_{\mu\nu}^m, \\
  && \label{psieq} \tilde{\nab}^2\psi =0. \eeqn
Its energy density and pressure for $\psi$ are the same given by \be
\rho_{\psi}=\frac{\dot{\psi}^2}{2}=p_{\psi}, \ee which describe a
stiff matter with $W_\psi=1$. Here we obtain canonical forms for
$\rho_\psi$ and $p_\psi$, in comparison with non-canonical forms of
$\rho_{BD}$ (\ref{rhobd}) and $p_{BD}$ (\ref{prebd}) in the
BD-frame.  A continuity equation of
$\dot{\rho}_{\psi}+3H(\rho_{\psi}+p_{\psi})=0$ [unlike (\ref{BDC})]
leads to the $\psi$-scalar equation (\ref{psieq}) exactly as \be
\label{quin}\ddot{\psi}+3H \dot{\psi}=0. \ee Importantly, the
Bianchi identity leads to the two conservation  laws separately \be
\nabla^\mu T_{\mu\nu}^\psi=0 \to \ddot{\psi}+3H
\dot{\psi}=0,~~\nabla^\mu T^m_{\mu\nu}=0. \ee Hence, we need the EOS
\begin{equation} W_{ \psi}=\frac{\rho_{\psi}}{p_{\psi}},
\end{equation} whereas we do not need to introduce the effective EOS
$W^{\rm eff}_{\psi}$ like (\ref{effeos}) arisen from (\ref{BDC}).

In the absence of matter,  the BD scalar plays a role of kinetic
matter. This kinetic matter evolves  as the conservation law is
satisfied by itself \be \label{MC}\dot{\rho}_{\rm BD} + 3H(
\rho_{\rm BD} + p_{\rm BD} ) = 0 \ee whose equation  of state (EOS)
is defined  by \be \label{ES}
   W_{\rm BD} \equiv \frac{p_{\rm BD}}{\rho_{\rm BD}}.
\ee  The solution to the first Friedmann and BD scalar equations is
given by \be \label{realsol} a(t)=t^{\frac{3(\omg+1)\pm
\sqrt{3(2\omg+3)}}{3(3\omg+4)}},
~~\Phi(t)=t^{\frac{1\mp\sqrt{3(2\omg+3)}}{3\omg+4}}. \ee Plugging
the above into (\ref{ES}), one finds its EOS as \be \label{EOS}
   W^{\pm}_{\rm BD} =\frac{3(\omg+2)\pm 2\sqrt{3(2\omg+3)}}{3\omg}~~{\rm with}~ \omg \ge -\frac{3}{2}.
\ee In the limit  of $\omg \to 0$, $W^{+}_{\rm BD}\to 4/0$, while
$W^{-}_{\rm BD}\to \frac{1}{3}$.  Their behavior is shown in Fig. 1.
Here we choose $W_{\rm BD}=W^{-}_{\rm BD}$ as the EOS for the BD
kinetic-matter. The EOS bound is given by $-1/3 \le W_{\rm BD} \le
1$. If one requires the condition (\ref{BDC}) together with
$\rho_{m}=0$,  the only allowable solution  is the case saturating
the lower bound
\begin{equation}
\omg=-\frac{3}{2} \to W^-_{\rm BD}=-\frac{1}{3}\end{equation}
 which  corresponds the solution to the
conformal relativity: $a(t)\sim t,~\Phi \sim 1/t^2, \rho_{\rm
BD}\sim 1/a^2, p_{\rm BD}=-\rho_{\rm BD}/3$. This case
 gives a zero acceleration of $\ddot{a}= 0$.
  Consequently, the perfect fluid interpretation of
the BD scalar is valid only for  $\omg=-3/2$~\cite{KimHS}.

When the CDM is present, we have to solve the different equations.
In the FRW spacetime,  equations take the forms
\begin{eqnarray}
&& H^2  = \frac{8\pi G}{3}\Big(\rho_{\rm BD} +
  \frac{\rho_{\rm m}}{\Phi}\Big), \label{hubble-bd}\\
&& \dot H = -4 \pi G \Big( \rho_{\rm BD}
  + p_{\rm BD} + \frac{\rho_{\rm m}}{\Phi}+\frac{p_{\rm m}}{\Phi}\Big), \label{hubble1-bd} \\
&& \ddot\Phi + 3H \dot\Phi  = \frac{8 \pi G}{2\omega_{\rm BD} +3}
\left ( \rho_m - 3 p_m \right ) , \label{field-bd}
\end{eqnarray}
where $\rho_{\rm m}$ is the CDM  density given by \be \rho_{\rm m} =
\frac{\rho_{\rm m}^0}{ a^{3}}\ee with the present dark matter
density $\rho_{\rm m}^0$.  It is convenient to use new variables as
\begin{equation}
\label{def-variable}
x = \ln a, ~~ \varphi = \frac{\Phi'}{\Phi}, ~~
\lambda = -\frac{H'}{H}
\end{equation}
where $'$ denote the derivatie with respect to $x$. Also we define
the density parameters
\begin{equation}
\Omega_{\rm BD} \equiv \frac{8 \pi G}{3H^2} \rho_{\rm BD}, ~~
\Omega_{\rm m} \equiv \frac{8 \pi G}{3H^2} \frac{\rho_{\rm m}}{\Phi} .
\end{equation}
Using the relations
\begin{eqnarray}
\label{rel1}\frac{\dot\Phi}{\Phi} &=& \frac{dx}{dt} \frac{d\Phi}{dx} \frac{1}{\Phi} = H \varphi, \\
\label{rel2} \frac{\ddot\Phi}{\Phi} &=& H^2 \left ( \varphi' +
\varphi^2 - \lambda \varphi \right ) ,
\end{eqnarray}
energy density and pressure are given, respectively, by
\begin{eqnarray}
\rho_{\rm BD} &=& \frac{H^2}{16\pi G} \left [
\omega_{\rm BD} \varphi^2
- 6 \varphi \right ], \\
p_{\rm BD} &=& \frac{H^2}{16\pi G} \left [ \omega_{\rm BD} \varphi^2
+ 4 \varphi -2 \lambda \varphi + 2 \left( \varphi' + \varphi^2
\right ) \right ].
\end{eqnarray}
The Bianchi identity (\ref{BDC}) takes into account  the energy
transfer between BD field and CDM,  while the CDM  evolves according
to its own conservation law
\begin{equation}
\dot\rho_{\rm m} + 3H \left ( \rho_{\rm m} +p_{\rm m} \right ) = 0.
\end{equation}
Eqs.(\ref{hubble-bd}), (\ref{hubble1-bd}) and (\ref{field-bd}) can
be written as
\begin{eqnarray}
&& 1 = \Omega_{\rm BD} + \Omega_{\rm m}, \label{r_hubble-bd}\\
&& \lambda = \frac{3}{2} + \frac{4 \pi G}{H^2} p_{\rm BD}, \label{r_hubble1-bd} \\
&&  \varphi' - \lambda \varphi + 3 \varphi + \varphi^2 =
    \frac{3}{2\omega_{\rm BD} +3} (1-\Omega_{\rm BD}), \label{r_field-bd}
\end{eqnarray}
where we  used the pressureless condition of $p_{\rm m} = 0$ for the
CDM. Solving Eq.(\ref{r_field-bd}) for $\varphi'$ and inserting it
into $p_{\rm BD}$ leads to
\begin{equation}
p_{\rm BD} = \frac{H^2}{16\pi G} \Big[
  \omega_{\rm BD} \varphi^2 -2 \varphi
  + \frac{6}{2\omega_{\rm BD}+3}(1-\Omega_{\rm BD})
  \Big].
\end{equation}
Substituting this  into Eq. (\ref{r_hubble1-bd}), we find
\begin{equation}
\label{lambda_def-bd} \lambda = \frac{3}{2} + \frac{1}{4} \Big[
  \omega_{\rm BD} \varphi^2 -2 \varphi
  + \frac{6}{2\omega_{\rm BD}+3}(1-\Omega_{\rm BD})
  \Big].
\end{equation}
A further relation  is found to be
\begin{equation}
\label{omega_bd-bd}
\Omega_{\rm BD} = \frac{1}{6} (\omega_{\rm BD} \varphi^2 -6 \varphi ) .
\end{equation}
Eq. (\ref{r_field-bd}) can be rewritten as
\begin{equation}
\varphi' = -\varphi^2 -3 \varphi
   + \frac{3(1-\Omega_{\rm BD})}{2\omega_{\rm BD} +3}  +\lambda \varphi . \label{phiprime-bd}
\end{equation}
Let us plug  $\lambda$ and $\Omega_{\rm BD}$  into Eqs.
(\ref{phiprime-bd}) and solve it numerically with the initial
condition.  On the other hand, we obtain the native  EOS for the BD
fluid \begin{eqnarray} W_{\rm BD} &=& \frac{p_{\rm BD}}{\rho_{\rm
BD}} = \frac{ \omega_{\rm BD} \varphi^2 -2 \varphi
  + \frac{6}{2\omega_{\rm BD}+3}(1-\Omega_{\rm BD})
}
{\omega_{\rm BD} \varphi^2 - 6\varphi}.
\end{eqnarray}
Considering  Eq. (\ref{BDC}) as
\begin{equation}
\dot\rho_{\rm BD} + 3 H \left ( 1 + W_{\rm BD}^{\rm eff} \right ) \rho_{\rm BD} = 0,
\end{equation}
we  obtain the effective EOS
\begin{eqnarray}
W_{\rm BD}^{\rm eff} &=& W_{\rm BD} - \frac{\rho_{\rm m}}{3\rho_{\rm BD}\Phi}\varphi \nonumber \\
\label{effeos} &=& \frac{\varphi}{3} + \frac{  \omega_{\rm BD}
\varphi^2 -4 \varphi
  + \frac{6}{2\omega_{\rm BD}+3}(1-\Omega_{\rm BD})
} {\omega_{\rm BD} \varphi^2 - 6\varphi}.
\end{eqnarray}
The initial value for
$\varphi$ is determined  by
\begin{equation}
\varphi(0)=\frac{3\pm\sqrt{3(2\Omega^{0}_{\rm BD}\omega_{\rm
BD}+3)}}{\omega_{\rm BD}}
\end{equation}
with $\Omega_{\rm BD}(0)=\Omega_{\rm BD}^0$. Here $+(-)$ sign
correspond to increasing (decreasing) $\Omega_{\rm BD}$ at $x=0$.
For $-$ signature, its evolution induces a singularity. This
behavior could be expected from  the critical points obtained by
solving the equation of $\varphi'= 0$.  The result is summarized in
the Table \ref{table_BD}.
\begin{table}
  \begin{center}
  \begin{tabular}{ccccc}
    \hline
    class & $\varphi$ &  $\Omega_{\rm BD}$ &
    $W_{\rm BD}^{\rm eff} $ &
    $W_{\rm BD}$ \\
    \hline

    (a) &
    $\frac{1}{\omega_{\rm BD}+1}$ &
    $-\frac{5\omega_{\rm BD}+6}{6(\omega_{\rm BD}+1)^2}$ &
    $\frac{1}{3(\omega_{\rm BD}+1)}$ &
    $-\frac{2(\omega_{\rm BD}+1)}{5\omega_{\rm BD}+6}$ \\

    (b) &
    $\frac{3+\sqrt{3(2\omega_{\rm BD}+3)}}{\omega_{\rm BD}}$ &
    1 &
    $W^+_{\rm BD}$ &
    $W^+_{\rm BD}$ \\

    (c) &
    $\frac{3-\sqrt{3(2\omega_{\rm BD}+3)}}{\omega_{\rm BD}}$ &
    1 &
    $W^-_{\rm BD}$ &
    $W^-_{\rm BD}$ \\
    \hline
  \end{tabular}
  \end{center}
\caption{List of critical points without  BD potential.}
\label{table_BD}
\end{table}
These critical points indicate asymptotic behaviors in the far
future and  far past.  In order to  test whether each critical point
is or not stable, we need to observe  the signature of $d \varphi' /
d\varphi$. If the signature is negative (positiven), it may be stable
for the far future evolution (far past evolution).  The viable
parameter range  for class (a) is  found by requiring the condition
of  $0 \le \Omega_{\rm BD} \le 1$ as
\begin{equation}
\omega_{\rm BD} < -\frac{3}{2}, ~~ -\frac{4}{3} \le \omega_{\rm BD} \le -\frac{6}{5}.
\end{equation}
However, if we demand the positive-definite energy density for the
BD fluid $\rho_{\rm BD} > 0$ and the negative-definite pressure
$p_{\rm BD} < 0$~\cite{KimHS}, the relevant range is determined
solely by
\begin{equation}
\omega_{\rm BD} < -\frac{3}{2}.
\end{equation}
In this case, the native and effective equations of state take the
bounds
\begin{equation}
W_{\rm BD},~W_{\rm BD}^{\rm eff}  > -\frac{2}{3}
\end{equation}
which means that the BD fluid without potential does not explain the
future phantom divide.
 For classes (b) and (c), these are nothing new  because we have the condition
\begin{equation}
\omega_{\rm BD} \ge -\frac{3}{2} \to W_{\rm BD}=W^{\rm eff}_{\rm BD}\ge
-\frac{1}{3}
\end{equation}
which corresponds to  the absence of the CDM  as is shown in
Eq.(\ref{EOS}).

Therefore, we note  that the role of BD scalar without potential
(equivalently, k-essence with non-canonical kinetic term only) as a
source generating the accelerating universe is very restricted
because it can at most describe ``$W_{\rm BD}(W_{\rm BD}^{\rm
eff})=-2/3$ acceleration" in the presence of the CDM. In the
presence of matters~\cite{KimHS}, the BD scalar $\Phi$ appears to
interpolate smoothly between the matter-dominated and accelerating
eras  by speeding up the expansion rate of the matter-dominated era
like ($a(t) \sim t^{2/3} \to
t^\alpha(\alpha={2(\omg+1)/(3\omg+4)}>2/3))$, while slowing down
that of accelerating phase derived by cosmological constant to some
degree like ($a(t) \sim e^{\bar{\chi}t} \to (1+\chi t)^{(2\omg
+1)/2}$). Hence, we have to include an appropriate potential to
obtain the phantom divide of $W_{\rm BD}(W^{\rm eff}_{\rm BD})=-1$.

\section{ BD cosmology with a potential}
\label{constant_alpha}

 The action for generalized BD theory is given
by\cite{Kim05}
\begin{equation}
\label{BS-action}
S = \int d^4 x \sqrt{-g} \left [
\frac{1}{16\pi G} \left (
\Phi R - \omega_{\rm BD} \frac{\nabla_\alpha \Phi \nabla^\alpha \Phi}{\Phi}
 - 16 \pi G U(\Phi)
\right )
\right ] + S_{\rm m},
\end{equation}
where $S_{\rm m}=\int d^4 x \sqrt{-g} {\cal L}_m$ is the action for
the other matter of the perfect fluid type and $U(\Phi)$ is a
potential for the BD scalar $\Phi$.  The equation of motions can be
obtained as
\begin{eqnarray}
G_{\mu\nu} =R_{\mu\nu} -\frac{1}{2}g_{\mu\nu} R
&=& \frac{\omega_{\rm BD}}{\Phi^2} \left \{
    \nabla_\mu\Phi\nabla_\nu\Phi -\frac{1}{2} g_{\mu\nu} \left (\nabla\Phi \right )^2
       \right \} \nonumber \\
&&+
  \frac{1}{\Phi} \left \{
    \nabla_\mu\nabla_\nu\Phi -g_{\mu\nu} \nabla^2\Phi - g_{\mu\nu} 8 \pi G U(\Phi)
   \right \} + \frac{8 \pi G}{\Phi} T^{\rm m}_{\mu\nu} \nonumber \\
&\equiv& 8 \pi G T^{\rm BD}_{\mu\nu} + \frac{8 \pi G}{\Phi} T^{
m}_{\mu\nu} ,
\end{eqnarray}
where $T^{\rm BD}_{\mu\nu}$ is  given by
\begin{equation}
T^{\rm BD}_{\mu\nu}=\frac{1}{8 \pi G} \left [
\frac{\omega_{\rm BD}}{\Phi^2} \left \{
    \nabla_\mu\Phi\nabla_\nu\Phi -\frac{1}{2} g_{\mu\nu} \left (\nabla\Phi \right )^2
       \right \} +
\frac{1}{\Phi} \left \{
    \nabla_\mu\nabla_\nu\Phi -g_{\mu\nu} \nabla^2\Phi - g_{\mu\nu} 8 \pi G U(\Phi)
   \right \}
  \right ]
\end{equation}
and $T^{ m}_{\mu\nu}$ is the energy-momentum tensor (\ref{emten}).
Equation for $\Phi$ is changed to be
\begin{equation}
\nabla^2\Phi + \frac{16 \pi G}{2\omega_{\rm BD}+3} \left (
   2 U(\Phi) - \Phi \frac{dU}{d\Phi}
  \right ) = \frac{8 \pi G}{2\omega_{\rm BD}+3} T^{\rm m
  \alpha}~_{\alpha}.
\end{equation}
 In
the context of dark energy, it is possible to construct a single
scalar model (scalar-tensor theory) based on the BD theory with
constant $\eta=-d\ln U(\phi)/d\phi$. This depends on the choice of
the potential $U(\Phi)$.

Setting $\Phi=F(\varphi)=e^{-2Q\phi}$, the exponential potential
$U_0e^{-\eta \phi}$ takes the power-law form
\begin{equation} \label{p-pot}
U(\Phi) = U_0 \Phi^{\alpha }
\end{equation}
with the constant $\alpha=\eta/2Q$. In the FRW spacetime
(\ref{frw}), three equations take the forms
\begin{eqnarray}
&& H^2  = \frac{8\pi G}{3}\Big(\rho_{\rm BD} +
  \frac{\rho_{\rm m}}{\Phi}\Big), \label{hubble}\\
&& \dot H = -4 \pi G \Big(\rho_{\rm BD} -  \frac{\rho_{\rm m}}{\Phi}
   - p_{\rm BD}-\frac{p_m}{\Phi}\Big), \label{hubble1} \\
&& \ddot\Phi + 3H \dot\Phi - \frac{16 \pi G}{2\omega_{\rm BD}+3} \left (
   2 U(\Phi) - \Phi \frac{dU}{d\Phi}\right ) = \frac{8 \pi G}{2\omega_{\rm BD} +3}
(\rho_m-3p_m). \label{field}
\end{eqnarray}
These equations are consistent with Ref.\cite{MH} but are slightly
different from Ref.\cite{Fsa}.  Regarding the BD field as a perfect
fluid, its energy and pressure are changed as~\cite{KimHS,KLM}
\begin{eqnarray} \label{noncr}
\rho_{\rm BD} &=& \frac{1}{16\pi G} \left [
\omega_{\rm BD} \left ( \frac{\dot\Phi}{\Phi}\right )^2
- 6 H \frac{\dot\Phi}{\Phi} + 16 \pi G \frac{U(\Phi)}{\Phi}
\right ], \\
\label{noncp}p_{\rm BD} &=& \frac{1}{16\pi G} \left [ \omega_{\rm
BD} \left ( \frac{\dot\Phi}{\Phi}\right )^2 + 4 H
\frac{\dot\Phi}{\Phi} + 2 \frac{\ddot\Phi}{\Phi} - 16 \pi G
\frac{U(\Phi)}{\Phi} \right ].
\end{eqnarray}
The Bianchi identity leads to the same relation as in
Eq.(\ref{BDC}).

In order to solve Eqs. (\ref{hubble}), (\ref{hubble1}) and
(\ref{field}), it is convenient to introduce new variables as
\begin{equation}
\psi = \frac{8 \pi G U(\Phi)}{H^2 \Phi},
\end{equation}
Using (\ref{rel1}) and (\ref{rel2}),  energy density and pressure
are expressed in terms of $\varphi$ and $\psi$, respectively, by
\begin{eqnarray} \label{den11}
\rho_{\rm BD} &=& \frac{H^2}{16\pi G} \left [
\omega_{\rm BD} \varphi^2
- 6 \varphi + 2\psi
\right ], \\
\label{pre11}p_{\rm BD} &=& \frac{H^2}{16\pi G} \left [ \omega_{\rm
BD} \varphi^2 + 4 \varphi -2 \lambda \varphi + 2 \left( \varphi' +
\varphi^2 \right ) - 2\psi \right ].
\end{eqnarray}
Then, Eqs. (\ref{hubble}), (\ref{hubble1}) and (\ref{field}) can be
written as
\begin{eqnarray}
&&  \Omega_{\rm BD} + \Omega_{\rm m}=1, \label{r_hubble}\\
&& \lambda = \frac{3}{2} + \frac{4 \pi G}{H^2} p_{\rm BD}, \label{r_hubble1} \\
&&  \varphi' - \lambda \varphi + 3 \varphi +
    \varphi^2 - \frac{2}{2\omega_{\rm BD}+3}
    \left ( 2 - \alpha \right ) \psi  =
    \frac{3}{2\omega_{\rm BD} +3} (1-\Omega_{\rm BD}). \label{r_field}
\end{eqnarray}
Solving Eq. (\ref{r_field}) for $\varphi'$, and inserting it into
$p_{\rm BD}$, we obtain the pressure
\begin{equation} \label{pre12}
p_{\rm BD} = \frac{H^2}{16\pi G} \Big[
  \omega_{\rm BD} \varphi^2 -2 \varphi +
   \frac{4(2-\alpha)}{2\omega_{\rm BD}+3} \psi
  + \frac{6}{2\omega_{\rm BD}+3}(1-\Omega_{\rm BD}) - 2 \psi
  \Big].
\end{equation}
Substituting this  into Eq. (\ref{r_hubble1}), we arrive at
\begin{equation}
\label{lambda_def} \lambda = \frac{3}{2} + \frac{1}{4} \Big[
  \omega_{\rm BD} \varphi^2 -2 \varphi +
   \frac{4(2-\alpha)}{2\omega_{\rm BD}+3} \psi
  + \frac{6}{2\omega_{\rm BD}+3}(1-\Omega_{\rm BD}) - 2 \psi
  \Big].
\end{equation}
Importantly, we  note that $\Omega_{\rm BD}$ can be written as
\begin{equation}
\label{omega_bd} \Omega_{\rm BD} = \frac{1}{6} (\omega_{\rm BD}
\varphi^2 -6 \varphi + 2 \psi ).
\end{equation}
Plugging this into Eq. (\ref{r_field}) leads to
\begin{equation}
\varphi' = -\varphi^2 -3 \varphi +
          \frac{2( 2 - \alpha)}{2\omega_{\rm BD}+3}\psi
   + \frac{3(1-\Omega_{\rm BD})}{2\omega_{\rm BD} +3}  +
     \lambda \varphi . \label{phiprime}
\end{equation}
On the other hand, from the definition of $\psi$, we obtain a newly
differential equation
\begin{equation}
\label{psiprime}
\psi' = \left( \alpha\varphi - \varphi+2\lambda \right ) \psi .
\end{equation}
Now we have to solve two coupled equations  (\ref{phiprime}) and
(\ref{psiprime}) for $\varphi$ and $\psi$ numerically with initial
conditions.

\begin{figure}[t!]
\centering
\includegraphics[width=0.8\textwidth]{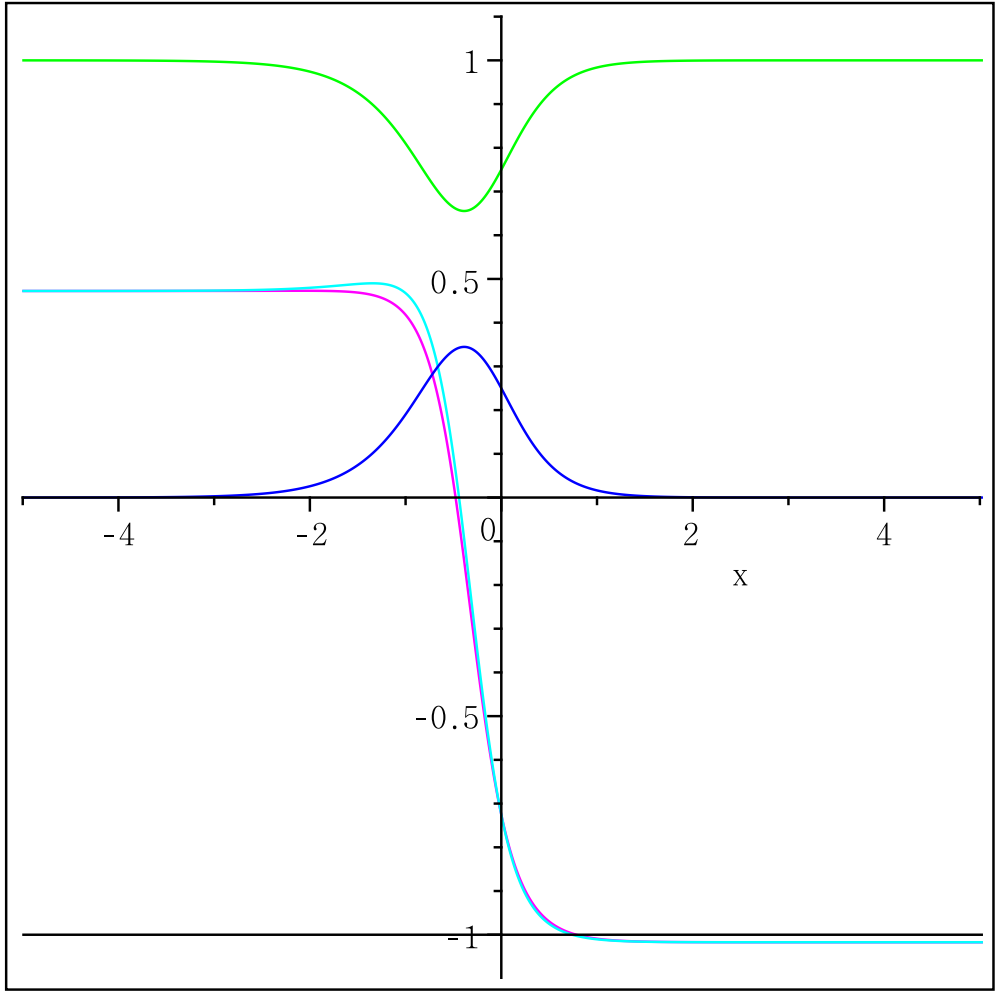}
\caption{Time evolution of the BD cosmology with
 potential $U=U_0\Phi^\alpha$  with $\omg=0$: $W_{\rm BD}$(magenta), $W^{\rm
eff}_{\rm BD}$(cyan), $\Omega_{\rm m}$(blue), $\Omega_{\rm
BD}$(green). The initial condition at $x=0$ is imposed by
$\alpha=1.449$, $\psi(0)=2.253$, $\varphi(0)=\frac{2.253}{3}-0.75=0.001$
(corresponding to $\Omega_{\rm BD}^0=0.75$), and $\Phi(0)=1.0$.}
\label{fig1}
\end{figure}

Considering (\ref{den11}) and (\ref{pre12}), we obtain the native
EOS  for BD field with potential
\begin{eqnarray} \label{eosn}
W_{\rm BD} &=& \frac{p_{\rm BD}}{\rho_{\rm BD}} = \frac{
\omega_{\rm BD} \varphi^2 -2 \varphi +
   \frac{4(2-\alpha)}{2\omega_{\rm BD}+3} \psi
  + \frac{6}{2\omega_{\rm BD}+3}(1-\Omega_{\rm BD}) - 2 \psi
}
{\omega_{\rm BD} \varphi^2 - 6\varphi +2\psi}.
\end{eqnarray}
This might not be  suitable for representing the true equation of
state for the BD scalar because of the non-conservation of this
fluid (\ref{BDC}).
 We remind the reader that (\ref{BDC}) could be rewritten as
\begin{equation}
\dot\rho_{\rm BD} + 3 H \left ( 1 + W_{\rm BD}^{\rm eff} \right ) \rho_{\rm BD} = 0,
\end{equation}
which implies an effective EOS
\begin{eqnarray} \label{eose}
W_{\rm BD}^{\rm eff} &=& W_{\rm BD} - \frac{\rho_{\rm m}}{3\rho_{\rm BD}\Phi}\varphi \nonumber \\
&=& \frac{\varphi}{3} + \frac{  \omega_{\rm BD} \varphi^2 - 4 \varphi +
   \frac{4(2-\alpha)}{2\omega_{\rm BD}+3} \psi
  + \frac{6}{2\omega_{\rm BD}+3}(1-\Omega_{\rm BD}) - 2 \psi
}
{\omega_{\rm BD} \varphi^2 - 6\varphi +2\psi} .
\end{eqnarray}
A typical solution is given in Fig. \ref{fig1} for $\omg=0$, showing
that $\Omega_{\rm BD}+\Omega_{m}=1$.  We observe that there exists a
phantom divide ($W_{\rm BD}=-1,~W_{\rm BD}^{\rm eff}=-1)$ as
confirmed by effective EOS  $W_{\rm BD}^{\rm eff}$.

In deriving this numerical solution, it was necessary to impose the
initial condition at $a_0=1(x=0)$ as an input of the current
observation data. It is impossible to construct  nearly past,
present, and future acceleration phases without fixing the initial
condition with the current observation data. Usually, we need one
initial condition to solve the first order differential equation.
Most of cases are needed to specify $t=0$ as the initial condition.
However, since we do not know the origin of the dark energy clearly,
we could not use $t=0$ as the initial condition to solve
(\ref{phiprime}) and (\ref{psiprime}) for  the accelerating phase.

 Since
the effective gravitational constant $G^{\rm eff} = G/\Phi$ varies
with time, it should satisfy the observed limits defined as
\begin{equation}
\frac{\dot G^{\rm eff}}{G^{\rm eff}}=-\frac{\dot \Phi}{\Phi} = -H \varphi \leq 10^{-13} {\rm yr}^{-1} .
\end{equation}
It  means that
\begin{equation}
H_0 \varphi(0) \leq 10^{-13} {\rm yr}^{-1},
\end{equation}
and, finally
\begin{equation}
\varphi(0) \leq 0.0013.
\end{equation}
Here we used the present Hubble parameter of  $H_0 = 77 $km/s/Mpc  =
$2.5 \times 10^{-18} {\rm s}^{-1} = 7.88 \times 10^{-11} {\rm
yr}^{-1}$.

Considering the density parameter of BD field as
\begin{equation}
\Omega_{\rm BD} \equiv \frac{\rho_{\rm BD}}{\rho_{\rm c}} =
\frac{\omega_{\rm BD} \varphi^2 -6 \varphi + 2 \psi}{6},~~\rho_{\rm
c} \equiv  \frac{8 \pi G}{3 H^2},
\end{equation}
$\Omega_{\rm BD}$ is regarded  as dark energy density
parameter and its current value will be  determined by observation as
$\Omega^0_{\rm BD}$. Hence two initial values of $\varphi$ and $\psi$ are not
independent, but they are related as
\begin{equation}
\omega_{\rm BD} \varphi^2 -6 \varphi(0) + 2 \psi(0) -6 \Omega^0_{\rm BD} = 0,
\end{equation}
which gives us
\begin{equation}
\varphi(0) = \frac{3 \pm \sqrt{9-2\omega_{\rm BD} (\psi(0)-\Omega^0_{\rm BD})}}{\omega_{\rm BD}}.
\end{equation}
For $\omega_{\rm BD} \rightarrow 0$, we can choose  $-$ sign so that
\begin{equation}
\varphi(0)=\frac{\psi(0)}{3} - \Omega^0_{\rm BD}.
\end{equation}

In order to find the asymptotic values for variable, we need to
determine the critical points from Eqs.(\ref{phiprime}) and
(\ref{psiprime}).   We list the critical points and corresponding
physical variables in Table \ref{table_fixed} and Table \ref{table_fixed0} for
$\omega_{\rm BD}=0$.
\begin{table}
  \begin{center}
  \begin{tabular}{cccccc}
    \hline
    class & $\varphi$ & $\psi$ & $\Omega_{\rm BD}$ &
    $W_{\rm BD}^{\rm eff} $ &
    $W_{\rm BD}$ \\
    \hline

    (a) &
    $\frac{1}{\omega_{\rm BD}+1}$ &
    0 &
    $-\frac{5\omega_{\rm BD}+6}{6(\omega_{\rm BD}+1)^2}$ &
    $\frac{1}{3(\omega_{\rm BD}+1)}$ &
    $-\frac{2(\omega_{\rm BD}+1)}{5\omega_{\rm BD}+6}$ \\

    (b) &
    $\frac{3+\sqrt{6\omega_{\rm BD}+9}}{\omega_{\rm BD}}$ &
    0 &
    1 &
    $W^+_{\rm BD}$ &
    $W^+_{\rm BD}$ \\

    (c) &
    $\frac{3-\sqrt{6\omega_{\rm BD}+9}}{\omega_{\rm BD}}$ &
    0 &
    1 &
    $W^-_{\rm BD}$ &
    $W^-_{\rm BD}$ \\

    (d) &
    $-\frac{3}{\alpha}$ &
    $\frac{3(\alpha +3 \omega_{\rm BD} +3)}{2\alpha^2}$ &
    $\frac{7\alpha + 6\omega_{\rm BD}+3}{2\alpha^2}$ &
    $-\frac{1}{\alpha}$ &
    $-\frac{2\alpha}{7\alpha + 6\omega_{\rm BD}+3}$ \\

    (e) &
    $-\frac{2(\alpha-2)}{\alpha+2\omega_{\rm BD}+1}$ &
    $\frac{(2\omega_{\rm BD}+3)\{ 6\omega_{\rm BD} - (\alpha-5)(\alpha+1)  \}}
      {(\alpha+2\omega_{\rm BD}+1)^2}$ &
    1 &
    $\frac{2\alpha^2-9\alpha-6\omega_{\rm BD}+1}{3(\alpha + 2\omega_{\rm BD}+1)}$ &
    $\frac{2\alpha^2-9\alpha-6\omega_{\rm BD}+1}{3(\alpha + 2\omega_{\rm BD}+1)}$ \\
    \hline
  \end{tabular}
  \end{center}
\caption{List of critical points with the power-law potential
(\ref{p-pot}).} \label{table_fixed}
\end{table}

\begin{table}
  \begin{center}
  \begin{tabular}{cccccc}
    \hline
    class & $\varphi$ & $\psi$ & $\Omega_{\rm BD}$ &
    $W_{\rm BD}^{\rm eff} $ &
    $W_{\rm BD}$ \\
    \hline

    (a) &
    1 &
    0 &
    $-1$ &
    $\frac{1}{3}$ &
    $-\frac{1}{3}$ \\

    (b) &
    N/A &
     &
     &
     &
     \\

    (c) &
    $-1$ &
    0 &
    1 &
    $\frac{1}{3}$ &
    $\frac{1}{3}$ \\

    (d) &
    $-\frac{3}{\alpha}$ &
    $\frac{3(\alpha +3)}{2\alpha^2}$ &
    $\frac{7\alpha + 3}{2\alpha^2}$ &
    $-\frac{1}{\alpha}$ &
    $-\frac{2\alpha}{7\alpha + 3}$ \\

    (e) &
    $-\frac{2(\alpha-2)}{\alpha+1}$ &
    $-\frac{3(\alpha-5)}
      {(\alpha+1)}$ &
    1 &
    $\frac{2\alpha^2-9\alpha+1}{3(\alpha + 1)}$ &
    $\frac{2\alpha^2-9\alpha+1}{3(\alpha + 1)}$ \\
    \hline
  \end{tabular}
  \end{center}
\caption{List of critical points in the presence of BD potential
(\ref{p-pot}) with $\omega_{\rm BD}=0$.} \label{table_fixed0}
\end{table}
\begin{figure}[t!]
\centering
\includegraphics[width=0.8\textwidth]{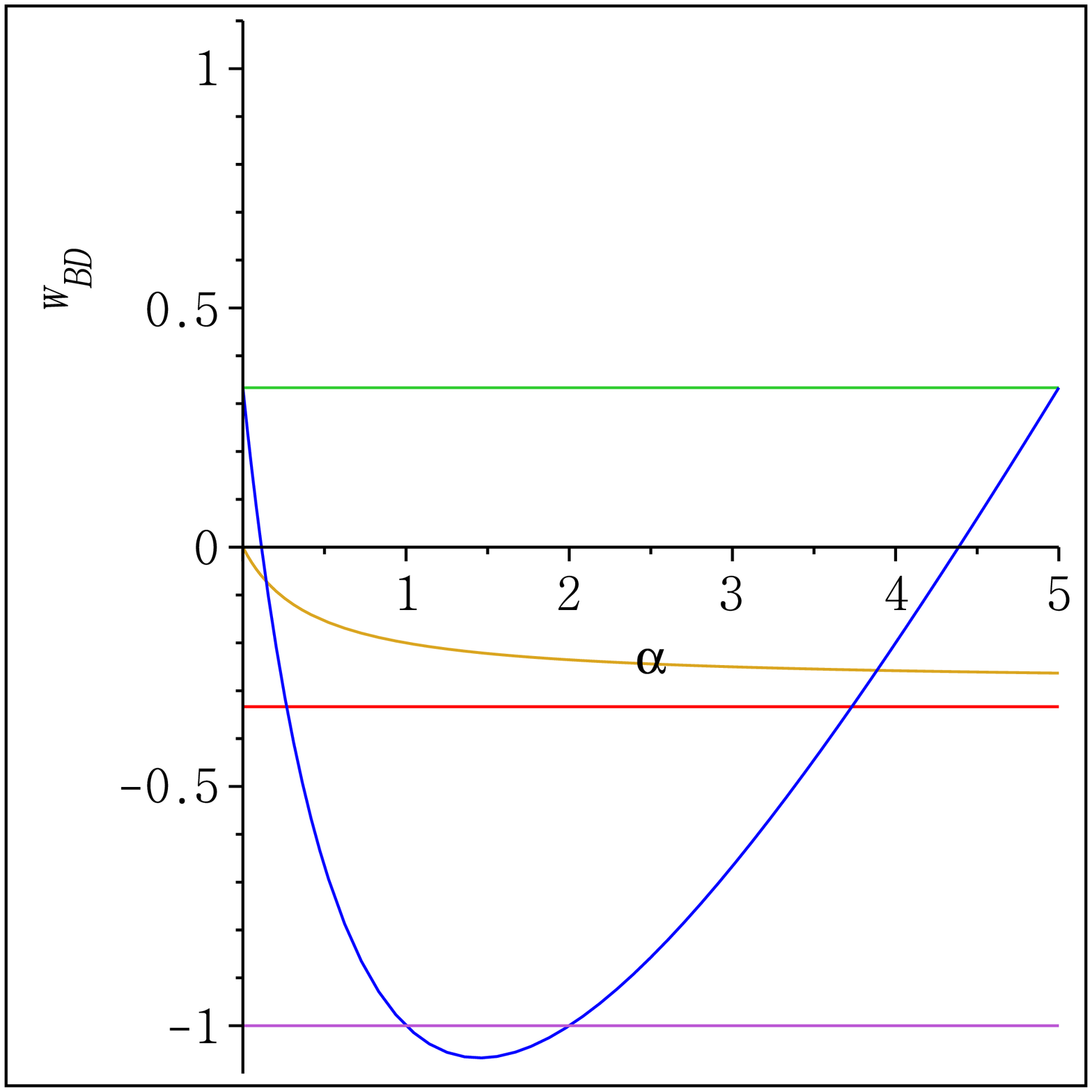}
\caption{The graphs of $W_{\rm BD}$ as a function of $\alpha$ for
the case of $\omega_{\rm BD}=0$. Class (a): red; (b): N/A; (c):
green; (d): dark yellow ; (e): blue.} \label{fig-w}
\end{figure}
\begin{figure}[t!]
\centering
\includegraphics[width=0.8\textwidth]{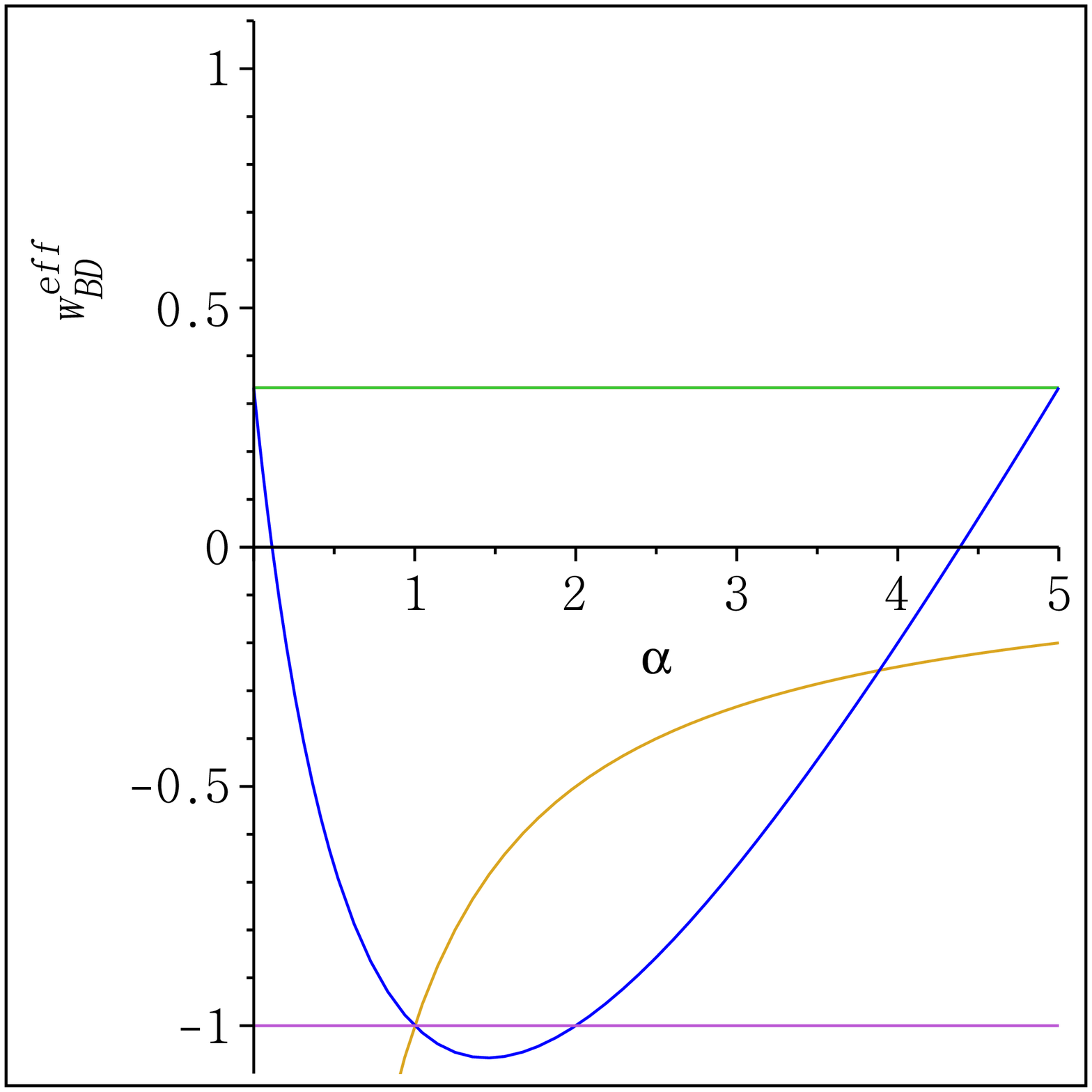}
\caption{The graphs of $W^{\rm eff}_{\rm BD}$ as a function of
$\alpha$ for the case of $\omega_{\rm BD}=0$.  Class (a): red (it is
overlapped with green); (b): N/A; (c): green; (d): dark yellow ;
(e): blue. } \label{fig2}
\end{figure}
Now we wish to analyze which  critical point is stable with time
evolution.  To determine the stability, we consider the perturbation
around the critical points. If the coefficients of the perturbation
is negative (positive), then it may be stable for future evolution
(past evolution). Actually, classes (a), (b) and (c) are nothing new
because these are exactly the same classes  in the absence of  the
potential (see Table \ref{table_BD}).  The class (d) is less
interesting because both its native and effective equations of state
do not provide the phantom divide.

The graph of equation of state $W_{\rm BD}$ as a function of
$\alpha$ is shown  in Fig. \ref{fig-w} with $\omg=0$.
 The graph of
effective equation of state $W^{\rm eff}_{\rm BD}$ as a function of
$\alpha$ is given in Fig. \ref{fig2} with $\omg=0$. The class (b) is
not available for $\omg=0$, as was mentioned by the BD
kinetic-matter using $W^+_{\rm BD}$.

 The class (e) is a newly interesting case. Let us study the class (e) more.  The solution to
the phantom divide of $W^{\rm eff}_{\rm BD}(\alpha)=-1$ is given by
\begin{equation}
\alpha_1 = 1, ~~~~ \alpha_2 = 2,
\end{equation}
while the solution to the dust matter of  $W^{\rm eff}_{\rm
BD}(\alpha)=0$ takes the forms
\begin{equation}
\alpha_3 = \frac{9-\sqrt{73}}{4}\simeq 0.114, ~~~~
\alpha_4 = \frac{9+\sqrt{73}}{4}\simeq 4.386.
\end{equation}
The solutions to the radiation of  $W^{\rm eff}_{\rm
BD}(\alpha)=\frac{1}{3}$ is given by
\begin{equation}
\alpha_5 = 0, ~~~~
\alpha_6 = 5.
\end{equation}
The minimum value of $W^{\rm eff}_{\rm BD}(\alpha)$
\begin{equation}
W^{\rm eff}_{\rm BD}(\alpha)\vert_{\rm min} =
4 \sqrt{\frac{2}{3}} - \frac{13}{3} \simeq -1.067,
\end{equation}
appears at
\begin{equation}
\alpha_{\rm min} = -1+\sqrt{6} \simeq 1.449.
\end{equation}

\begin{figure}[t!]
\centering
\includegraphics[width=0.8\textwidth]{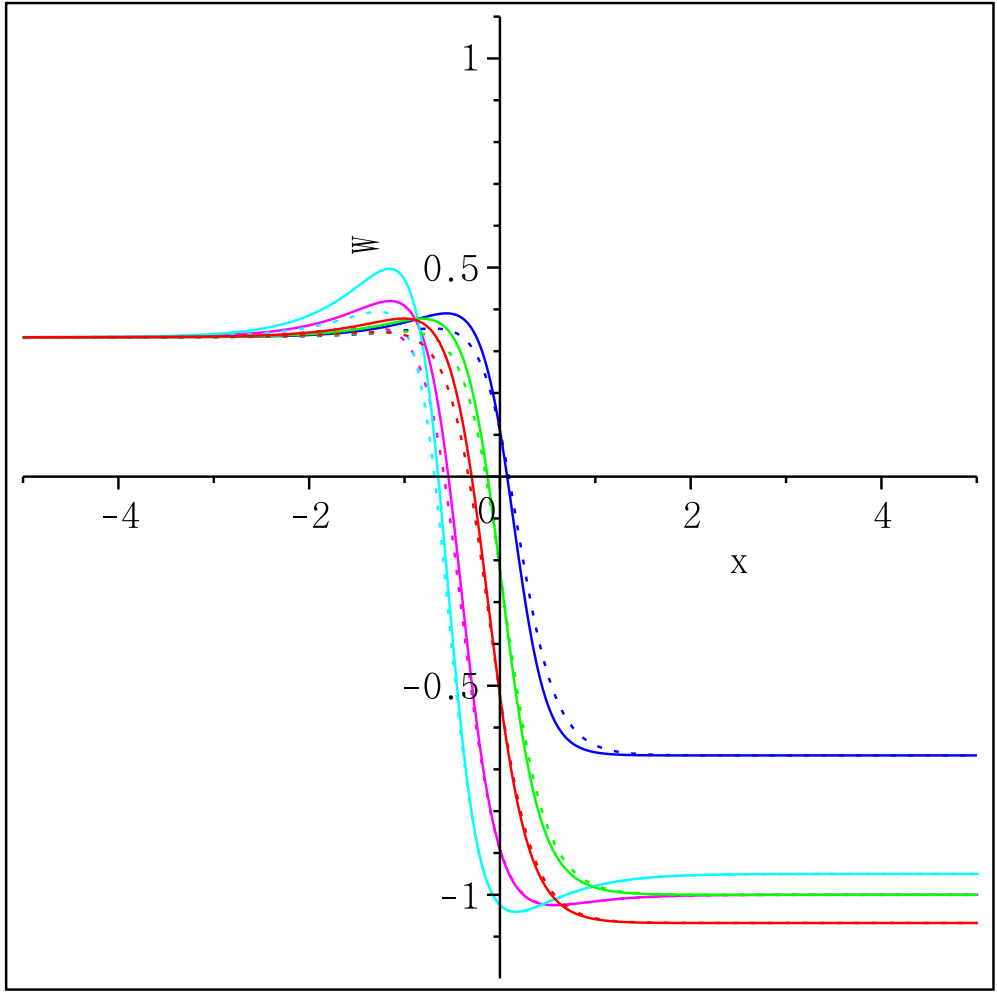}
\caption{Time evolution of  $W_{\rm BD}$(dotted curve) and $W_{\rm
BD}^{\rm eff}$(solid curve) for $\alpha=0.500$ (blue),
$\alpha=1.000$ (green), $\alpha=1.449$ (red as in Fig. \ref{fig1}),
$\alpha=2.000$ (magenta), $\alpha=2.200$ (cyan). } \label{fig_w}
\end{figure}

Finally, we mention the $\alpha$-dependent evolutions of two
equations of state $W_{\rm BD}$ and $W^{\rm eff}_{\rm BD}$. As  is
shown in Fig. \ref{fig_w}, for $\alpha  \le 1$, there is no phantom
divide, while for $\alpha>1$, there is  phantom divide. For
$\alpha>\alpha_{\rm min}$, there are two crossings of $W^{\rm
eff}_{\rm BD}=-1$. For $1<\alpha<\alpha_{\rm min}$, there is one
crossing of $W^{\rm eff}_{\rm BD}=-1$ and $ W^{\rm eff}_{\rm BD}$
approaches de Sitter value of $-1$ for $\alpha=1,2$.

\section{BD cosmology as $f(R)$ gravity}
The crossing of the phantom divide could be understood in the viable
$f(R)$ gravity models~\cite{f-models}. Recently, it was shown that
the viable four $f(R)$ models generally exhibit the crossing of the
phantom divide in the future evolution when using the EOS $w_{\rm
DE}=p^f_{\rm DE}/\rho^f_{\rm DE}$~\cite{BGL}. Hence, it is very
important to see whether the future crossing of phantom divide is
available for the BD cosmology with the corresponding potential.
Hence, we analyze the BD cosmology with the potential
\begin{equation} \label{potential}
U(\Phi)=U_0 \left \{ 1 - C (1-\Phi)^p \right \},
\end{equation}
which could be obtained by considering the equivalence between the
$f(R)$ gravity and the scalar-tensor theory (BD theory  with
potential and $\omg=0$)~\cite{TUMTY}. Actually, this potential is
an approximation to Hu-Sawicki and Starobinsky $f(R)$ models. Here
two parameters $C$ and $p$ are chosen
\begin{equation}
 0<C<1,~~~0<p<1.
 \end{equation}
Previous analysis could be applied to here  when defining  $\alpha$
to be
\begin{equation}
\frac{dU(\Phi)}{d\Phi}\equiv \alpha \frac{U(\Phi)}{\Phi},
\end{equation}
where $\alpha$ is determined  by
\begin{equation}
\alpha(\Phi) = Cp\frac{\Phi(1-\Phi)^{p-1}}{1-C(1-\Phi)^p} .
\end{equation}
Then, we could  use the same equations in Sec. \ref{constant_alpha}.
Since $\alpha$ is a function of $\Phi$, we  need to solve the
differential equation for $\Phi$.  Hence the full equations to be
solved  are given by
\begin{eqnarray}
\varphi' &=& -\varphi^2 -3 \varphi + \frac{2}{2\omega_{\rm BD}+3}
    \left ( 2 - \alpha(\Phi) \right ) \psi
   + \frac{3}{2\omega_{\rm BD} +3} (1-\Omega_{\rm BD}) +\lambda \varphi, \\
\psi' &=& \left( \alpha(\Phi) \varphi - \varphi+2\lambda \right ) \psi, \\
\Phi' &=& \varphi \Phi .
\end{eqnarray}
Relevant variables take the forms
\begin{eqnarray}
\lambda &=& \frac{3}{2} + \frac{1}{4} \Big[
  \omega_{\rm BD} \varphi^2 -2 \varphi +
   \frac{4(2-\alpha(\Phi))}{2\omega_{\rm BD}+3} \psi
  + \frac{6}{2\omega_{\rm BD}+3}(1-\Omega_{\rm BD}) - 2 \psi
  \Big], \\
\Omega_{\rm BD} &=& \frac{1}{6} (\omega_{\rm BD} \varphi^2 -6 \varphi + 2 \psi ), \\
W_{\rm BD} &=& \frac{ \omega_{\rm BD} \varphi^2 -2 \varphi +
   \frac{4(2-\alpha(\Phi))}{2\omega_{\rm BD}+3} \psi
  + \frac{6}{2\omega_{\rm BD}+3}(1-\Omega_{\rm BD}) - 2 \psi
}
{\omega_{\rm BD} \varphi^2 - 6\varphi +2\psi}, \\
W_{\rm BD}^{\rm eff}
&=& \frac{\varphi}{3} + \frac{  \omega_{\rm BD}
\varphi^2 -4 \varphi +
   \frac{4(2-\alpha(\Phi))}{2\omega_{\rm BD}+3} \psi
  + \frac{6}{2\omega_{\rm BD}+3}(1-\Omega_{\rm BD}) - 2 \psi
}
{\omega_{\rm BD} \varphi^2 - 6\varphi +2\psi} .
\end{eqnarray}

\begin{table}
  \begin{center}
  \begin{tabular}{ccccccc}
    \hline
    class & $\varphi$ & $\psi$ & $\Phi$ & $\Omega_{\rm BD}$ &
    $W_{\rm BD}^{\rm eff} $ &
    $W_{\rm BD}$ \\
    \hline

    (a) &
    $0$ &
    $3$ &
    $1-e^Z$ &
    $ 1$ &
    $\frac{1}{3} -\frac{2}{3}\frac{Cp\Phi(1-\Phi)^{p-1}}{1-C(1-\Phi)^p}$ &
    $\frac{1}{3} -\frac{2}{3}\frac{Cp\Phi(1-\Phi)^{p-1}}{1-C(1-\Phi)^p}$ \\

    (b) &
    $1$ &
    $0$ &
    $0$ &
    $-1$ &
    $\frac{1}{3}$ &
    $-\frac{1}{3}$ \\

    (c) &
    $-1$ &
    $0$ &
    $0$ &
    $1$ &
    $\frac{1}{3}$ &
    $\frac{1}{3}$ \\

    (d) &
    $4$ &
    $15$ &
    $0$ &
    $1$ &
    $\frac{1}{3}$ &
    $\frac{1}{3}$ \\

    \hline
  \end{tabular}
  \end{center}
\caption{List of critical points with the  potential
(\ref{potential}). Here $Z$ is the solution of
$Z(p-1)=\ln\frac{-2}{C[e^Z(p-2)-p]}$.} \label{table_fR}
\end{table}

\begin{figure}[t!]
\centering
\includegraphics[width=0.8\textwidth]{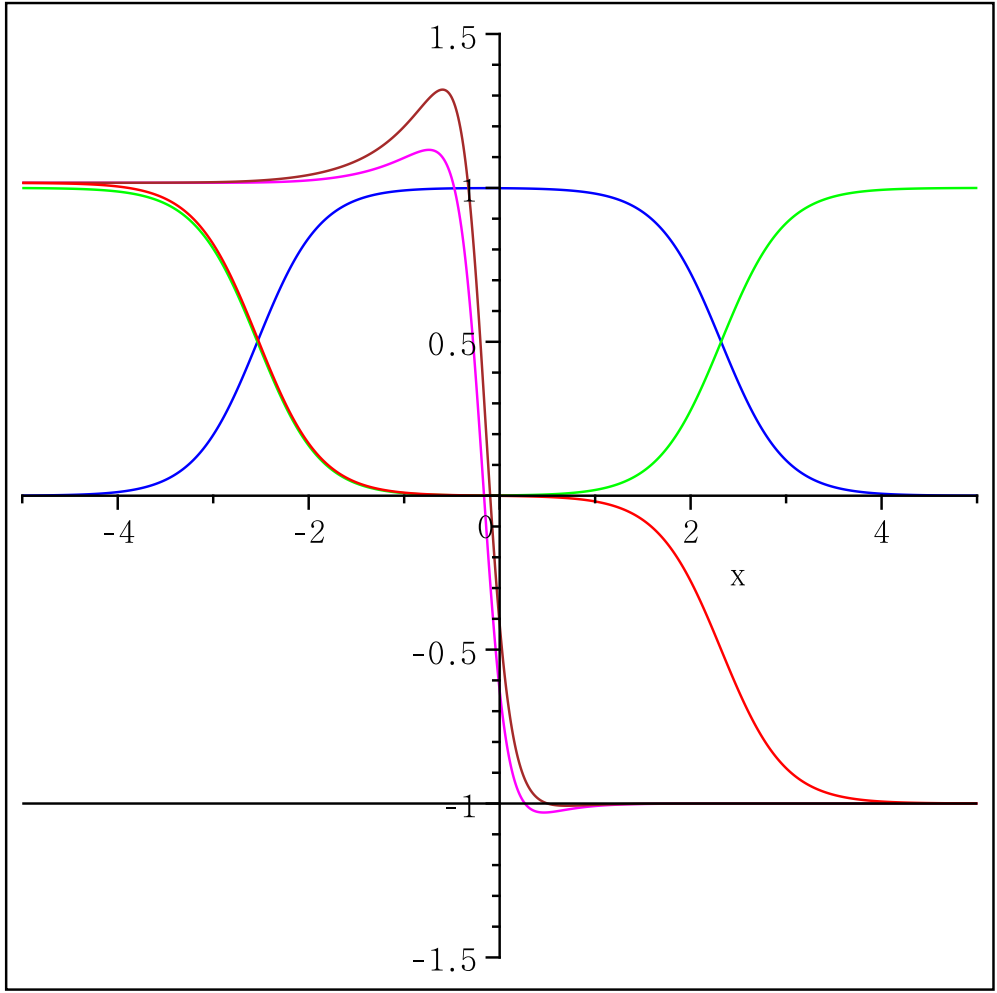}
\caption{The graphs of $\Omega_{\rm m}$(blue), $\Omega_{\rm
BD}$(green), $W^{\rm eff}_{\rm BD}$(magenta), $W_{\rm BD}$(brown),
$w_{\rm eff}$(red: definition used in Ref.\cite{TUMTY}) as a
function of $x$ with $\omg=9998.5$. The imposed condition is
$C=0.7$, $Q=0.01$, $p=0.2$, $\varphi(0) = 6.47\times 10^{-4}$,
$\psi = 2.85\times 10^{-3}$(corresponding to $\Omega_{\rm
BD}(0)=10^{-3}$, $\Phi(0) = 5.0\times 10^{-13}$).}
\label{fig-check}
\end{figure}

In order to check  that our system is working properly,  we first
recover Tsujikawa et al's result~\cite{TUMTY} by taking
$\omega_{\rm BD}=9998.5(Q=0.01)$.   We have recovered their result
of fig. 1 correctly,  which  is shown  in Fig. \ref{fig-check}.
This figure shows that for $0<C<1$, the matter-dominated phase
with $w_{\rm eff} \simeq 0$ is followed by the  de Sitter phase
with $w_{\rm eff} \simeq -1$.  In contrast to this, our equations
of states $W_{\rm BD}$ and $W_{\rm BD}^{\rm eff}$ show that the BD
field evolves from a stiff matter with $W_{\rm BD}(W_{\rm BD}^{\rm
eff})=1$ in the far past  to de Sitter phase with  $W_{\rm
BD}(W_{\rm BD}^{\rm eff})=-1$ in the far future. We note that the
phantom divide appears in the near future.  Also, we wish to point
out that  the initial condition used in~\cite{TUMTY} do not
provide a proper evolution. This means that the density parameter
$\Omega_{\rm BD}$ of BD field becomes negative, showing  an
unphysical case.  This explains why equations of sate diverge at
some points.   However, our modified initial condition gives a
correct evolution for all relevant physical variables.  Our
initial condition is chosen by requiring the nonnegative density
parameter of $\Omega_{\rm BD}(x_{\rm min}) \ge 0$ on  whole
evolution.  In Figure \ref{fig-check},  we have used $x_{\rm
min}=0$ and $\Omega_{\rm BD}(x_{\rm min}) = 10^{-3}$. We note that
the definition of energy density and pressure for the BD are
different from those in~\cite{TUMTY},  but the definition of
density parameter $\Omega_{\rm BD}$ is the same. Furthermore, they
have analyzed  the evolution only for the positive $x \ge 0$
(future direction), as one can see from the Figure
\ref{fig-check}. However, the general tendency of its whole
evolution is almost the same as in the power-law  potential $U=U_0
\Phi^\alpha$.

Furthermore,
 we have to mention that two equations of
state $W_{\rm BD}$ and $W_{\rm BD}^{\rm eff}$ defined by
Eq.(\ref{eosn}) and (\ref{eose}) are slightly   different from the
notation used in ~\cite{TUMTY}
\begin{equation}
w_{\rm eff}=-1-\frac{2}{3} \frac{\dot{H}}{H^2}
\end{equation}
which was  defined from the total conservation law
\begin{equation}
\dot{\rho}_{\rm tot}+3H(\rho_{\rm tot}+p_{\rm tot})=0 \to
\dot{\rho}_{\rm tot}+3H\rho_{\rm tot}(1+w_{\rm eff})=0,~~\rho_{\rm
tot}=\rho_{m}+\tilde{\rho}_{\rm BD}
\end{equation}
with $\tilde{\rho}_{\rm BD}=\Phi \rho_{\rm BD}$. Hence $w_{\rm eff}$
represents the equation of state for the whole matters in the
universe.

The asymptotic critical points are listed in Table \ref{table_fR}.
 Actually, tow classes (b) and  (c)  are nothing new
because these are nearly the same classes  in the absence of  the
potential (see classes (a) and (c) in Table \ref{table_fixed0}).
The class (d) is less interesting because it is similar to class
(c)  in Table \ref{table_fixed0}). We note that all of these
belong to unstable critical points. On the other hand, the class
(a) represents the asymptote of  $f(R)$ gravity models and its
critical point indicates the par future  behavior, showing that
$W_{\rm BD}(W^{\rm eff}_{\rm BD}) \to -1$.

\begin{figure}[t!]
\centering
\includegraphics[width=0.8\textwidth]{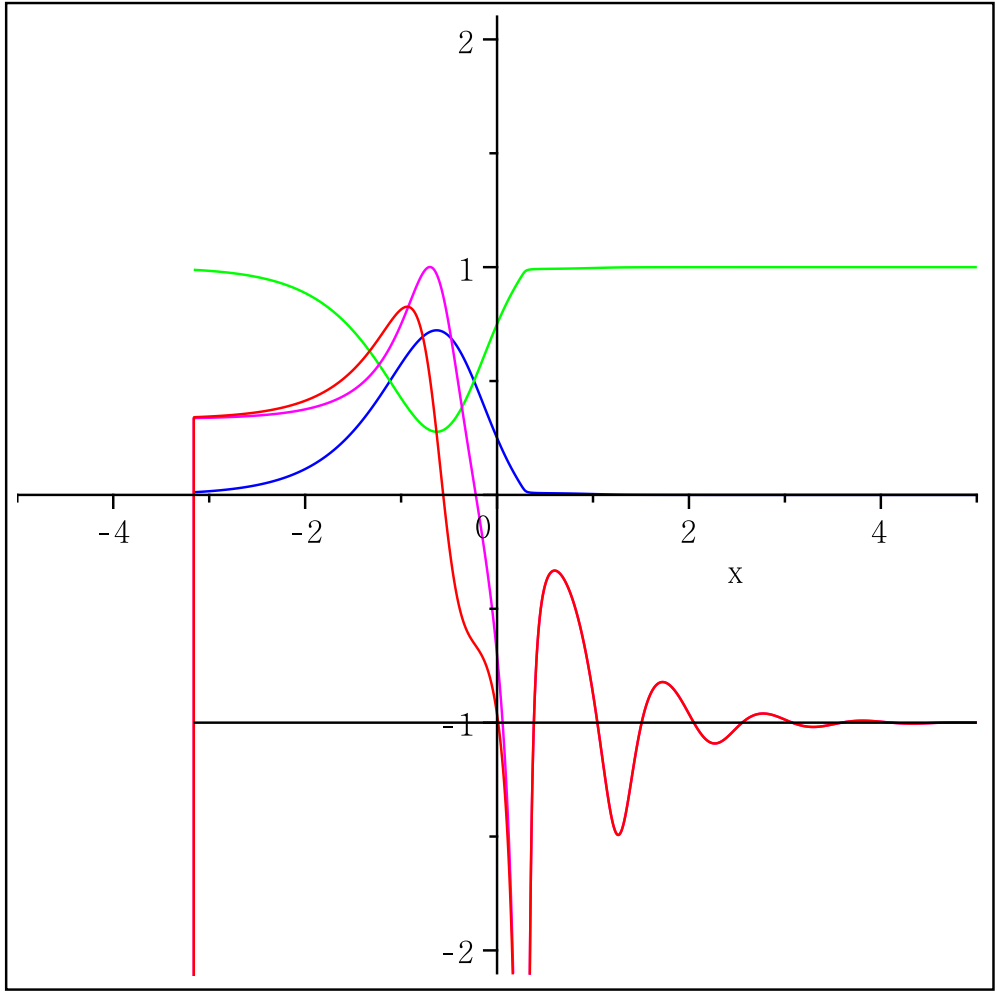}
\caption{A  time evolution of BD fluid  with potential
(\ref{potential}) and  $\omega_{\rm BD}=0$ ($f(R)$-gravity): $W_{\rm
BD}$(magenta), $W^{\rm eff}_{\rm BD}$(red), $\Omega_{\rm m}$(blue),
$\Omega_{\rm BD}$(green). The initial condition at $x=0$ is imposed
as  $C=0.82$, $p=0.09$, $\varphi(0)=2.35$, $\psi(0)=9.3$
(corresponding to $\Omega_{\rm BD}^0=0.75$), $\Phi(0)=0.6$.  A
singularity appears at $x=-3.16$ when evolving toward the far past,
while there is a damped oscillatory evolution toward the far future.
} \label{fig-evolve}
\end{figure}

 In order to have the
correspondence between $f(R)$ gravity and the BD theory, we have
to choose the BD parameter to be zero ($\omega_{\rm BD}=0$). In
this case, it is not easy  to impose appropriate initial
conditions of $\Omega_{\rm m}=0.25$ and $\Omega_{\rm BD}=0.75$ by
adjusting   two potential parameters $C$ and $p$.  A
 time evolution is depicted in Fig. \ref{fig-evolve}.  As  this figure is shown,
 there is a singularity at a past of $x=-3.16$ which reflects that the mapping is problematic. In this case, however, the
 equation of state $W_{\rm BD}$ approaches $-1$ (de Sitter spacetime) oscillatory as
 the universe evolves toward the far future. Also, the effective  equation of state $W_{\rm BD}^{\rm eff}$
 does show the nearly same  behavior toward
 the far future.  This confirm the presence of future crossing of phantom divide
 which appeared in the four viable $f(R)$ models~\cite{BGL} clearly.

\section{Discussions}
We have employed the BD cosmology to explain  the accelerating
universe and future crossing of phantom divide. In this work, we
regard the BD field as a perfect fluid model. First, the role of BD
scalar without potential (k-essence with non-canonical kinetic term
only) as a source generating the accelerating universe is very
restricted because it could describe ``$W_{\rm BD}(W^{\rm eff}_{\rm
BD})=-2/3$ acceleration" in the presence of the CDM.

Turning on the power-law potential (\ref{p-pot}), the BD cosmology
could describe the accelerating universe in the interaction with the
CDM. In this case, we have used both the equation of state $W_{\rm
BD}$ and effective equation of state $W_{\rm BD}^{\rm eff}$ to check
whether the phantom phase appears.  Explicitly, the BD field acts as
a radiation field in the far past, whereas it plays a role of
phantom field in the far future.  This is compared with the case
without the potential where the BD field acts as a radiation field
in the far past, while it plays a role of an accelerating matter in
the far future.   This shows that the presence of the potential is
crucial for  obtaining a phantom divide in the BD cosmology.

Concerning the BD description of $f(R)$-gravity, we have chosen
the potential in Eq.(\ref{potential}) inspired by Hu-Sawicki and
Starobinsky $f(R)$ models.  For the case of $\omg=0$, the
evolution of $\Omega_{m}$ and $\Omega_{\rm BD}$ are similar to the
power-law potential (\ref{p-pot}), but there exists a singularity
at $x=-3.16$, which restricts evolving  toward the far past of
$x=-\infty$ after imposing the initial condition at $x=0$.
However, we have found that the universe evolves toward the far
future of $x=\infty$  nicely. Both native and effective equations
of state converge  to $-1$ (the de Sitter spacetime of class(a) in
Table \ref{table_fR}) oscillatory, which indicates that the BD
description is working for showing a future crossing of phantom
divide appeared in the viable $f(R)$ gravity models~\cite{BGL}.

\begin{acknowledgments}
This work was supported by the National Research Foundation of Korea
(NRF) grant funded by the Korea government(MEST) (No. 2010-0028080).
\end{acknowledgments}


\end{document}